\documentclass[letterpaper, 10 pt, conference]{ieeeconf}  

\IEEEoverridecommandlockouts                              

\usepackage[utf8]{inputenc}
\usepackage[T1]{fontenc}

\usepackage{amsmath,graphicx,subfigure,color,amssymb,makecell,multirow,booktabs, algorithm,algpseudocode,cite,placeins}
\usepackage[english]{babel}

\DeclareMathOperator*{\argmin}{argmin} 

\DeclareMathOperator{\vol}{Vol}

\begin{document}

\title{\LARGE \bf
 {Voronoi Shaping with Efficient Encoding}
}


\author{H. Buglia
         and R. R. Lopes
 \thanks{$^{1}$Henrique Buglia  was with Télécom ParisTech, Paris, France.
 He is now with the School of Electrical and Computer Engineering, University of Campinas, 13083-970 Campinas, SP, Brazil.
        {\tt\small henrique.buglia@gmail.com}}
        
 \thanks{$^{2}$Renato R. Lopes is with the School of Electrical and Computer Engineering, University of Campinas, CP 6101, 13083-970 Campinas, SP, Brazil
        {\tt\small rlopes@decom.fee.unicamp.br}}%
        }

\maketitle
\thispagestyle{empty}
\pagestyle{empty}

\begin{abstract}

In this letter, we propose a Voronoi shaping method with reduced encoding complexity. The method works for integer shaping and coding lattices satisfying the chain $\Lambda_s \subseteq \textbf{K}\mathbb{Z}^n \subseteq \Lambda_c$, with \textbf{K} an integer diagonal matrix. {This assumption is easily satisfied for lattices obtained from error-correcting codes. For those lattices, using this strategy, an explicit set of coset representatives is obtained and lattice encoding complexity is reduced to the linear code encoding complexity.} Our proposal {is illustrated in construction-D lattices with Gosset ($E_8$) and Leech ($\Lambda_{24}$) lattices as shaping lattices.}

\end{abstract}

\section{INTRODUCTION}

Lattices~\cite{conway2013sphere} provide a structured way to find dense sphere packings in high dimensions. As such, they are a natural tool for transmitting digital information over the additive white Gaussian noise (AWGN)~\cite{forney1998modulation} {channel}. In fact, they can be shown to achieve the capacity of this channel~\cite{zamir2014lattice}.

To transmit using lattice codes~\cite{forney1998modulation}  we must first find a lattice with a dense packing of points in an $n$-dimensional space. This is known as the coding lattice. Second, we must select only the points of the coding lattice inside a given region of the $n$-space. This operation is called shaping, and ensures that the power constraints are satisfied, and that a finite number of points can be transmitted.

For the first step, many authors have proposed good lattice designs based on several error-correcting codes~\cite{di2017leech,matsumine2018construction,yan2012construction,da2018multilevel}, yielding so-called construction-A or construction-D lattices. However, usually a simple hypercube is used for shaping. This can be implemented with very low complexity, but at the expense of the gains provided by good shaping.

{In Voronoi shaping \cite{khodaiemehr2016practical,liu2018construction}, only points of the coding lattice inside the Voronoi region of a shaping lattice are transmitted. These points are known as coset leaders of the quotient group}. Recently an efficient {nested} Voronoi shaping scheme was proposed in~\cite{di2017leech} for construction-A lattices. {The key is a low-complexity strategy to find coset representatives of the quotient group, not necessarily in the Voronoi region of the shaping lattice. Then, these representatives are quantized to the Voronoi region}. In this letter we generalize this method {to a broad class of lattices that includes those based on the construction D.}

{We illustrate} our proposal  { for the} construction-D lattices obtained from the extended BCH code in \cite{matsumine2018construction} and  {the SC-LDPC codes in \cite{vem2014multilevel}.} In these examples, shaping is performed, respectively, by the Gosset lattice $E_8$, {and the Leech lattice $\Lambda_{24}$}. {We also compare the complexity and performance of our scheme with QC-LDPC lattices proposed in \cite{khodaiemehr2016practical}}.

\section{PRELIMINARIES}

A lattice $\Lambda$ is defined as the discrete set

\begin{equation}\label{eq:Lattice_Definition}
\Lambda =   	\left \{ \textbf{x} = \textbf{G}\cdot \textbf{b}, \textbf{b} \in \mathbb{Z}^n  \right \}. 
\end{equation}
The $n \times n$ matrix $\textbf{G}$ is a lattice generator matrix.

{In this paper we assume integer lattices, for which the $\textbf{G}$ is full rank and has integer elements}.

The shortest-distance quantization of a point $\textbf{y} \in \mathbb{R}^n$ is the point of the lattice $\Lambda$ that is {the} closest to $\textbf{y}$:
\begin{equation}\label{eq:Quantization}
Q_{\Lambda}(\textbf{y}) = \argmin \limits_{\textbf{x} \in \Lambda}||\textbf{y}-\textbf{x}||^2.
\end{equation}
{A region $\mathcal{F} \subset \mathbb{R}^n$ is called fundamental region for a lattice $\Lambda$ if shifts of $\mathcal{F}$ by lattice points are disjoint and cover $\mathbb{R}^n$, that is, $\Lambda + \mathcal{F} = \mathbb{R}^n$. The volume of any fundamental region, also known as the volume of the lattice, $\vol(\Lambda)$, is equal to $\det(\textbf{G})$\cite{costa2017lattices}.} {An important fundamental region of $\Lambda$ is} the Voronoi region $\mathcal{V}$, {which} is the set of points that are closer to the origin than to any other point of the lattice $\Lambda$:
\begin{equation}\label{eq:Voronoi_Region}
\mathcal{V}(\Lambda) = \left \{ \textbf{y} \in \mathbb{R}^n \mid Q_{\Lambda}(\textbf{y}) = \textbf{0}  \right \}.
\end{equation}
{Other important fundamental regions are given in the following lemma:}

{\textbf{Lemma 1~\cite{kurkoski2018encoding}:} Let \textbf{G} be a triangular generator matrix of $\Lambda$}\footnote{ {A triangular generator matrix always exists for any full rank integer lattice, e.g, the Hermite normal form of \textbf{G} \cite{costa2017lattices}.}}, {with diagonal elements $g_{ii}$. Let \textbf{P} be a triangular matrix with diagonal elements $p_{ii} = g_{ii}$.}

{Then, the following parallelotope is a fundamental region of $\Lambda$:}
\begin{equation}\label{eq:fund_Region1}
{\mathcal{P}(\textbf{P}) =  \{
\textbf{Ps}
, 0 \leq s_i < 1, i = 1,\dots,n   \}} 
\end{equation}
\hfill $\square$

Two lattices are nested if $\Lambda_s \subseteq \Lambda_c$.
For any $\textbf{x} \in \Lambda_c$, the set $\textbf{x} + \Lambda_s$ is the \textit{coset} of $\Lambda_s$ in $\Lambda_c$ containing $\textbf{x}$, and \textbf{x} is called a coset representative. If $\textbf{x} \in {\cal V}(\Lambda_s)$, $\textbf{x}$ is called a coset leader.
The quotient group $\Lambda_c/\Lambda_s$ is the set of all cosets, i.e.,

\begin{equation}\label{eq:Quotient_Group}
\frac{\Lambda_c}{\Lambda_s} = \left \{ \textbf{x} + \Lambda_s \mid \textbf{x} \in \Lambda_c  \right \} .
\end{equation}

{\textbf{Lemma 2 \cite{forney2000sphere} }Consider the lattice partition chain $\Lambda_s \subseteq \Lambda_{r-1} \subseteq \dots \subseteq \Lambda_1 \subseteq \Lambda_c$. Let $A_1, \dots, A_r$ be sets of coset represenatives of the quotient groups $\Lambda_c/\Lambda_1,\dots,\Lambda_{r-1}/\Lambda_s$, respectively. Then, 

\eqref{eq:Quotient_Group}, can be rewritten as}
\begin{equation}\label{eq:Quotient_Group_chain}
{\frac{\Lambda_c}{\Lambda_s} = \left \{ \textbf{a}_1 + \dots + \textbf{a}_r + \Lambda_s \mid \textbf{a}_i \in A_i  \right \}}
\end{equation}

The number of distinct cosets, $M$, is the cardinality of the quotient group, and satisfies the following equation~\cite{costa2017lattices}:
\begin{equation}\label{eq:Quotient_Group_Card}
M = \left | \frac{\Lambda_c}{\Lambda_s} \right|= \frac{\vol{(\Lambda_s)}}{\vol{(\Lambda_c)}} = \left|\frac{\det{\textbf{G}_s}}{\det{\textbf{G}_c}}\right|
\end{equation}

A Voronoi constellation or lattice code $\mathcal{C}$ is the set of points of $\Lambda_c$ inside the Voronoi region of $\Lambda_s$: $\mathcal{C} = \Lambda_c \cap \mathcal{V}(\Lambda_s)$ \footnote{{Usually, a translation vector \textbf{d}, which ensures minimal transmit power, is applied, such that $\mathcal{C} = (\Lambda_c - \textbf{d}) \cap \mathcal{V}(\Lambda_s)$. Is this letter, this translation is omitted since it does not provide significant gains in the simulation results.}}. In this case, $\Lambda_s$ is called shaping lattice, $\Lambda_c$ is called coding lattice and $\mathcal{V}(\Lambda)$ is called the shaping region.

The power savings achieved by a given shaping region, when compared to a standard Pulse Amplitude Modulation (PAM) constellation, is captured by the shaping gain, defined in~\cite{forney1998modulation} \cite{forney1989multidimensional}.

\section{{ENCODING}}

In this section, we begin by proving a theorem that gives a simple method to {explicitly} compute a complete set of distinct  coset representatives of the quotient group $\Lambda_s/\Lambda_c$. The point in $\mathcal{C}$ to be transmitted can be determined by first choosing a coset representative $\textbf{x}$ using our theorem, then computing the element of the coset in $\mathcal{V}(\Lambda_s)$:
\begin{equation}\label{eq:Mudolo-s_Operation}
\textbf{x}' = \textbf{x} - Q_{\Lambda_s}(\textbf{x}).
\end{equation}
{\textbf{Theorem 1}: Let $\Lambda_s$ and $\Lambda_c$ be any integer lattices satisfying the lattice chain $\Lambda_s \subseteq \textbf{K}\mathbb{Z}^n \subseteq \Lambda_c$, for $\textbf{K}$ a diagonal matrix with entries $k_i$. Let a triangular generator matrix of $\Lambda_s$ be} 

\begin{equation}\label{eq:Hermite_normal_gamma}
{\textbf{G}_s =
\begin{pmatrix} 
    g_{s_{1,1}} & 0 & \dots & 0 \\
    g_{s_{2,1}} & g_{s_{2,2}} & \dots &  0\\
    \vdots & \ddots & \ddots &  \vdots\\
     g_{s_{n,1}} &  \dots &g_{s_{n,n-1}} & g_{s_{n,n}}\\ 
\end{pmatrix} 
\in \mathbb{Z}^{n \times n}}.
\end{equation}

{Define the set $\mathcal{S}$ as the Cartesian product:}

\begin{equation}\label{eq:set_S}
{
\mathcal{S} = \{0, \dots, \frac{g_{s_{1,1}}}{k_1}-1\} \times \dots \times \{0, \dots, \frac{g_{s_{n,n}}}{k_n}-1\}.} \footnote{{Note that, $g_{s_{1,1}}/k_i$ is an integer, since $\Lambda_s \subseteq \textbf{K}\mathbb{Z}^n$.}}
\end{equation}

{Let $\mathcal{X} = \Lambda_c \cap \mathcal{P}(\textbf{K})$, be the set of points of $\Lambda_c$ in the parallelotope associated with $\textbf{K}$, as defined in~\eqref{eq:fund_Region1}.}

{Then, a complete set of coset representatives of the quotient group $\Lambda_c/\Lambda_s$ is given by }

\begin{equation}\label{eq:Coset_general}
{\mathcal{X} +  \textbf{K}\cdot\mathcal{S} = 
\left \{ \textbf{x} + \textbf{K} \cdot\textbf{s} \mid \textbf{x} \in \mathcal{X}, \textbf{s} \in \mathcal{S}  \right\}}.
\end{equation}

{\textit{Proof}: A complete set of coset representatives of the quotient group $\Lambda_c/\Lambda_s$ is the set of points of $\Lambda_c$ inside any fundamental region of $\Lambda_s$ \cite{zamir2014lattice}. Likewise, a complete set of coset representatives of $\Lambda_c/\textbf{K}\mathbb{Z}^n$ is the set of points of $\Lambda_c$ inside the fundamental parallelotope of $\textbf{K}\mathbb{Z}^n$, which is a hyperrectangle $\mathcal{P}(\textbf{K})$, with sides $(k_1,\dots,k_n)$. This is the set $\mathcal{X}$. By lemma 1, a fundamental region of $\Lambda_s$ is the hyper-rectangle with sides $(g_{s_{1,1}},\dots,g_{s_{n,n}})$. Thus a complete set of coset representatives of $\textbf{K}\mathbb{Z}^n/\Lambda_s$ can be written as $\textbf{K}\mathcal{S}$ with $\mathcal{S}$ given in \eqref{eq:set_S} because of the triviality of the lattice $\textbf{K}\mathbb{Z}^n$. Finally, by lemma 2, a complete set of coset representatives of the quotient group $\Lambda_s/\Lambda_c$ is the sum of the coset representatives of the quotients $\Lambda_c/\textbf{K}\mathbb{Z}^n$ and $\textbf{K}\mathbb{Z}^n/\Lambda_s$, which is $\mathcal{X} + \textbf{K}\mathcal{S}$ as stated in \eqref{eq:Coset_general}.
\hfill $\square$
}

{Theorem 1 provides an efficient way to explicitly construct a set of coset representatives for any integer lattices satisfying the chain $\Lambda_s \subseteq \textbf{K}\mathbb{Z}^n \subseteq \Lambda_c$. This is direct for near-ellipsoidal lattice codes $\Lambda_c/\textbf{K}\Lambda_c$, with \textbf{K} an integer diagonal matrix~\cite{ragot2003near}. In this case, $\Lambda_s = \textbf{K}\Lambda_c  \subseteq \textbf{K}\mathbb{Z}^n \subseteq \Lambda_c$. Self-similar lattices codes $\Lambda_c/k\Lambda_c$ are a particular case of near-ellipsoidal lattices, with $\textbf{K} = k\textbf{I}$, so the theorem also applies direct to them. In general, for other lattices pairs, a shaping lattice $\Lambda_s \subseteq \textbf{K}\mathbb{Z}^n$ can be found by first choosing any integer lattice $\Lambda'$, then computing $\Lambda_s = \textbf{K}\Lambda'$. Since $\Lambda'$ is integer, this ensures that $\textbf{K}\Lambda' \subseteq \textbf{K}\mathbb{Z}^n$.}

{Theorem 1 has an interesting geometric interpretation. To determine a coset representatives, we first apply a hyperrectangular shaping to the coding lattice $\Lambda_c$, obtaining an element of $\mathcal{X}$. Then, we translate the resulting point by an element of the set $\textbf{K}\mathcal{S}$. The resulting coset representative lies inside the hyperrectangle with sides $(g_{s_{1,1}},\dots,g_{s_{n,n}})$.}

\section{{APPLICATION TO CONSTRUCTION-D LATTICES}}

{In this section, we apply theorem 1 to construction-D lattices. We start by showing how theorem 1 applies to these lattices. We then describe the encoding and indexing procedure, followed by the shaping lattice design, and end with a complexity analysis of our scheme. The strategy proposed in this section is a generalization to construction-D lattices of the strategy proposed in~\cite{di2017leech} for construction-A lattices.
} 

{\textbf{Corollary 1: Construction-D Lattices}: Let $C_0 \subseteq C_1 \subseteq \cdots \subseteq C_a = \mathbb{F}_q^n$ be a sequence of nested linear block code over $\mathbb{F}_q^n$ and let $\Lambda_c = \sum_{i=0}^{a-1} q^iC_i + q^a\mathbb{Z}^n$ be a construction-D coding lattice~\cite{conway2013sphere,matsumine2018construction}. Let $\Lambda'$ be any integer lattice with a triangular generator matrix \textbf{G}' with entries $g_{ij}'$. Choose $\Lambda_s = q^a \Lambda'$. A complete set of coset representatives of $\Lambda_c/\Lambda_s$ is:}
\begin{equation}\label{eq:Coset_Construction-D}
{\sum_{n=0}^{a-1} q^iC_i + q^a\mathcal{S} = 
\left \{ \sum_{n=0}^{a-1} q^i\textbf{c}_i + q^a\textbf{s} \mid \textbf{c}_i \in C_i, \textbf{s} \in \mathcal{S}  \right\}},
\end{equation}
{where}
\begin{equation}\label{eq:set_S_D}
\begin{split}
{\mathcal{S} = \{0, \dots, g'_{1,1} - 1\} \times \dots \times \{0, \dots, g'_{n,n} - 1\}.}
\end{split}
\end{equation}
 
{\textit{Proof:} The proof follows directly from theorem 1 by setting $\textbf{K} = q^a\textbf{I}$ and noting that, by construction,}
\begin{equation}\label{eq:Coset_Construction-D}
{
\mathcal{X} = \sum_{i=0}^{a-1}q^iC_i}
\end{equation}
\subsection{Encoding {and Indexing Procedure}}

{Choosing $\Lambda_c$, $\Lambda_s$ and $\textbf{K}$ as in corollary 1, encoding is done as follows.}

\begin{enumerate}
\item Select message vectors $\textbf{u}_i \in \mathbb{F}_q^{k_i}$, $i = 0, \ldots a - 1$.
\item Encode each message $\textbf{u}_i$ in order to obtain $\textbf{c}_i$.
\item Select an element $\textbf{s} \in \mathcal{S}$ and find the coset leader by computing $\textbf{x} = \sum_{i=0}^{a-1}q^i\textbf{c}_i + q^a\textbf{s}.$ 
\item Find the element of the coset of \textbf{x} inside the Voronoi region of the  shaping lattice $\Lambda_s: \textbf{x}' = \textbf{x} - Q_{\Lambda_s}(\textbf{x})$.

\end{enumerate}

Indexing, or {demapping}, is the reverse of encoding, producing the message $\textbf{m} = (\textbf{u}_0,\textbf{u}_1,\dots,\textbf{u}_{a-1},\textbf{s})$ given a Voronoi constellation point \textbf{x}'.  {The procedure to find each $\textbf{u}_i$ follows the indexing of multilevel lattices construction as in \cite{forney2000sphere}, and the procedure to find \textbf{s} is analogous to section III-B of \cite{di2017leech}.}

\subsection{{Shaping} Lattice Design}\label{sec:design}

To decrease quantization complexity, we can construct {$\Lambda'$} in corollary 1 as the direct sum of copies of an $n'$-dimensional lattice {with generator matrix $\textbf{G}_s'$. This construction retains the shaping gain and quantization complexity of the low dimensional lattice \cite{ferdinand2016low}}. If $n$ is a multiple of $n'$, the generator matrix of {$\Lambda'$} is written as \cite{zamir2014lattice}, 
\begin{equation}\label{eq:Gamma_generator_Matrix}
{\textbf{G}'  =
\begin{pmatrix} 
    \alpha \textbf{G}_s' & 0 & \dots & 0 \\
    0 & \alpha \textbf{G}_s' & \dots &  0\\
    \vdots & \ddots & \ddots &  \vdots\\
    0 &  \dots & 0 & \alpha \textbf{G}_s'\\ 
\end{pmatrix} 
\in \mathbb{Z}^{n \times n}},
\end{equation}
where {\textbf{G}'} is the matrix described in {corollary 1} and {$\textbf{G}_s'$} is the low dimensional shaping lattice generator matrix in {triangular form} and $\alpha$ is a scaling factor. 

Note that {$\textbf{G}_s'$}  is scaled by $\alpha$ for two reasons. First, it makes {$\textbf{G}'$}  an integer matrix and consequently {$\Lambda_s$} an integer lattice as required in theorem 1. Second it allows us to change the information rate, by expanding the shaping region and consequently increasing the number of points of $\mathcal{C}$. {Obviously, this expansion also increases the transmit power.} Note that $\Lambda_s$ and $\Lambda_c$ satisfy the requirements of theorem 1, because $\Lambda_s = {q^a\Lambda'} \subseteq q^a\mathbb{Z}^n \subseteq \Lambda_c \subseteq \mathbb{Z}^n$.

With this configuration, we can calculate the information rate $R$ of our transmission system. Using equation \eqref{eq:Quotient_Group_Card} and the fact that $R = \log_2(M)/n$, $R_c = \sum_{i=0}^{a-1} k_i/n$ and $\vol{(\Lambda_c)} = q^{n(a-R_c)}$, we have
\begin{equation}\label{eq:Information_Rate_shap}
\begin{split}
R =
\log_2\alpha + \frac{1}{n'}\log_2(\det{\textbf{G}_s'}) + R_c\log_2q\hspace{0.2cm} \text{bits/dim}.
\end{split}
\end{equation}

\subsection{{Complexity}}\label{sec:complexity}

{Voronoi shaping, as proposed in \cite{zamir2014lattice} and \cite{kurkoski2018encoding}, computes $\textbf{G}\cdot\textbf{b}$ to find the coset representative, followed by the quantization operation $Q(\textbf{x})$. If quantization is performed in small dimensions as proposed in section \ref{sec:design}, the overall encoding complexity is dominated by the matrix multiplication, whose complexity is generally $O{(n^2)}$.}

{In contrast, our proposal relies on theorem 1, which shows how the coset representatives can be explicitly computed. Thus, for coding lattices based on error-correcting codes, the complexity of this step is the same as the encoding complexity of the underlying code, which is usually smaller than $O{(n^2)}$.}

In \cite{khodaiemehr2016practical}, the authors use a construction-A lattice for shaping. The structure of the underlying code can be exploited for simpler, though sub-optimal, quantization. However, good coding lattices may not provide good shaping gains, especially in small dimensions. For instance, \cite{khodaiemehr2016practical} shows a shaping gain of 0.776 dB in dimension 60 using optimal quantization, and negative shaping gains when using sub-optimal quantization. In contrast, the Leech lattice provides a shaping gain of 1.03 dB in 24 dimensions \cite{forney1989multidimensional}, where optimal quantization is much simpler. Sub-optimal quantization enables the use of much larger dimensions. However, only negative coding gains are reported in \cite{khodaiemehr2016practical} for this type of quantization.

\section{SIMULATION RESULTS}

\subsection{Extended BCH Lattice Codes} \label{comp2}

 {The proposed scheme is first applied to the 128-dimensional 2-levels construction-D lattice obtained with two extended binary BCH codes as desined in \cite{matsumine2018construction}.}

In this example, $\textbf{G}_s'$ is the unimodular generator matrix of $E_8$. Its dimension is a divisor of the dimension of BCH codes, which enables the use of direct sums for building the shaping lattice. Also, in dimension eight, the $E_8$ lattice has the greatest packing density, is the best known quantizer \cite{conway2013sphere,costa2017lattices}, and its dimension is low enough that sphere decoding is feasible. The codes are the same as in \cite{matsumine2018construction}, and the ordered statistics (OSD) algorithm described in \cite{fossorier1995soft} is used for decoding.

\begin{figure}[h]
  \centering
	\includegraphics[width=8cm]{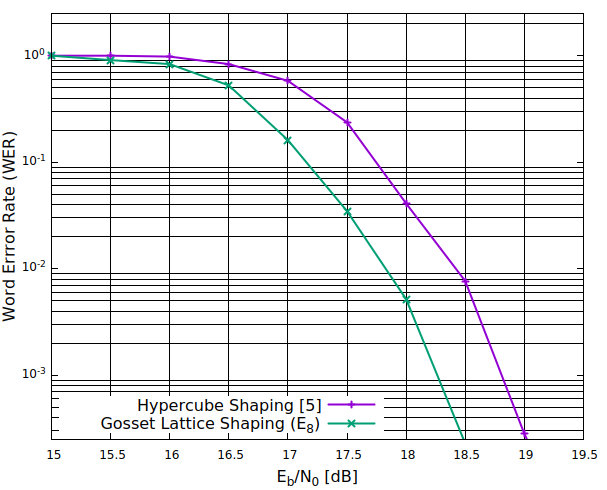}
  \caption{WER performance of 2-level extended BCH code lattices with
dimension n = 128 over AWGN channel constrained by cubic lattice and $E_8$ lattice.}
  \label{fig:EsN0}
\end{figure}

Figure \ref{fig:EsN0} shows the word-error rate (WER) as a function of $E_s/N_0$ for two shaping strategies: $E_8$ and hypercube shaping, achieved by choosing $\textbf{G}' = 2\textbf{I}_n$, which yields an 8-PAM constellation. This choice ensures that both shaping strategies achieve the same rate. We consider two-dimensional passband transmission, so the symbol energy is computed per two dimensions. In comparison with the hypercube shaping, a gain of 0.63 dB is obtained for WER of $10^{-3}$ when using the $E_8$ as shaping lattice. This value is close to the theoretical shaping gain for the $E_8$ lattice, 0.65\,dB~\cite{forney1998modulation}. This small difference is possibly to the several approximations involved in the definition and computation of the theoretical shaping gain \cite{forney1998modulation}.

\subsection{{SC-LDPC Lattice Codes}} \label{comp3}
{The proposed strategy was also applied to the construction-D lattice based on nested binary spatially-coupled low-density parity-check codes (SC-LDPC) proposed in \cite{vem2014multilevel}. We now use $\textbf{G}_s$' as the unimodular Leech $\Lambda_{24}$ generator matrix \cite{conway2013sphere,costa2017lattices}. It provides nominal shaping gain of 1.03\,dB~\cite{forney1989multidimensional}. In order to have an integer $\Lambda_{24}$ lattice, we set $\alpha = \sqrt{8}$ in \eqref{eq:Gamma_generator_Matrix} which using \eqref{eq:Information_Rate_shap} yields an additional rate of 1.5 bits/dim ($\log_2\sqrt{8}$), in comparison with a standard hypercube, obtained by setting $\textbf{G} = \textbf{I}_n$, yielding to a 8-PAM, or equivalently a 3 level construction with cubic shaping.}

\begin{figure}[h]
  \centering
	\includegraphics[width=8cm]{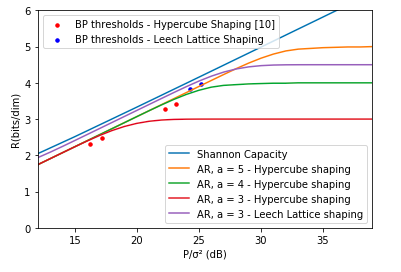}
  \caption{Achievable rates (AR) for different levels of construction-D lattices and with different shaping regions.}
  \label{fig:EsN0}
\end{figure}

{Figure 2 shows the achievable rates (sum-capacity of a 3, 4 and 5
level Construction-D lattice codes) of the construction proposed in \cite{vem2014multilevel} and the achievable rate of our scheme using the Leech lattice for 3 levels as a function of SNR. The SNR
required to successfully decode 10 consecutive codewords using message passing algorithm are shown as belief propagation (BP) thresholds for the same code designs in \cite{vem2014multilevel}. Our scheme is simulated for the case $R_c = 3.96$ and $Rc = 3,71$ where the additional 1.5 bits/dim is due the utilization of an integer leech lattice $\Lambda_{24}$.}

\section{CONCLUSIONS}

{A general method to identify an explicit set of cosets representatives and constructing Voronoi constellations was presented. The scheme is valid for all full rank lattices that satisfy theorem 1. It is shown that reduced encoding complexity is obtained for lattices from codes if an hypercube shaping is performed before standard shaping. It is also shown that the set of cosets representatives obtained lies inside an hyperrectangle. Our method was compared with QC-LDPC, SC-LDPC and BCH lattice codes. In all cases, reduced encoding complexities and shaping gains were obtained using Voronoi shaping scheme with good small dimensional lattices such as $\Lambda_{24}$ and $E_8$ lattices.}

\addtolength{\textheight}{-12cm}   




\section*{ACKNOWLEDGMENT}

This study was financed in part by the Coordenação de Aperfeiçoamento de Pessoal de Nivel Superior - Brasil (CAPES) - Finance Code 001 and by the National Council for Scientific and Technological Development, grant 305480/2018-9.

\bibliographystyle{IEEEbib}
\bibliography{main}

\end{document}